\documentclass[12pt] {article}      
\usepackage{amsmath,amssymb, amsthm, eucal,accents}  
\usepackage{latexsym, graphicx, amscd}
\usepackage{newtxtext,newtxmath}      
\usepackage[T1]{fontenc}     
\usepackage{enumerate}       
\usepackage{float}                 
\usepackage{mathtools}
\usepackage{tikz}
\usetikzlibrary{matrix, arrows, calc, fit}
\usetikzlibrary{arrows, decorations.markings}
\tikzstyle{AArrow} = [thick, decoration={markings,mark=at position 1 with {\arrow[semithick]{open triangle 60}}},%
   double distance=1.4pt, shorten >= 5.5pt,
   preaction = {decorate},
   postaction = {draw,line width=1.4pt, white,shorten >= 4.5pt}]
\tikzstyle{AArroww} = [semithick, white,line width=1.4pt, shorten >= 4.5pt]
\usepackage[strict]{changepage}
\usepackage{relsize}
\theoremstyle{plain}
\newtheorem{theorem}{Theorem}[section]            
\newtheorem{proposition}[theorem]{Proposition}  

\theoremstyle{definition}
\newtheorem{definition}[theorem]{Definition}

\numberwithin{theorem}{section}
\numberwithin{equation}{section}
\numberwithin{figure}{section}
\newcommand{\gaction}[2]{\genfrac{}{}{0.5pt}{}{#1}{#2}%
                        \!\lower2pt\hbox{\rotatebox[origin=c]{-90}{{$\looparrowright$}}}}

\newcommand{\dotfraction}[2]{\genfrac{}{}{0.5pt}{}{#1}{#2}%
                        \!\lower.5pt\hbox{{$\circ$}}}
\renewcommand\appendix{\par
  \setcounter{section}{0}
  \setcounter{subsection}{0}
  \setcounter{figure}{0}
  \setcounter{table}{0}
  \renewcommand\thesection{Appendix \Alph{section}}
  \renewcommand\thefigure{\Alph{section}\arabic{figure}}
  \renewcommand\thetable{\Alph{section}\arabic{table}}
}
\usepackage{titlesec}                                                           
\titlelabel{\thetitle.\ }                                                           
\titleformat*{\section}{\fontsize{14pt}{14pt} \bf}                   
\usepackage{fancyhdr}
\usepackage{sidecap}  
\usepackage[hang,small,sc]{caption}

\def\QED{ $\square$}

\def\ee{\varepsilon}
\def\ff{\varphi}

\def\moplus{\boxplus}
\def\voplus{\oplus}

\usepackage{xcolor}
\newcommand{\Dfrac}[2]{%
  \ooalign{%
    $\genfrac{}{}{1.7pt}0{#1}{#2}$\cr%
    $\color{white}\genfrac{}{}{.7pt}0{\phantom{#1}}{\phantom{#2}}$}%
}

\begin{document}

\title{\bf The loop of relativistic velocities \\
as a deformation of the menhir loop}
\author{Jerzy Kocik                  
\\ \small Department of Mathematics\\[-24pt]
\\ \small Southern Illinois University, Carbondale, IL62901
\\ \small jkocik{@}siu.edu  
}
\date{}

\maketitle


\begin{abstract}
\noindent
The algebra of the relativistic composition of velocities 
is shown to be isomorphic to an algebraic loop defined on division algebras.
This makes  calculations in special relativity effortless and straightforward,
unlike the the standard formulation, which consists of a rather convoluted algebraic equation.
The elegant  appearance of the new formula brings about an additional value.
\\[3pt]
{
\\[3pt]
}
\end{abstract}


\def\disk{\overset{\circ}}
\def\diskk{\accentset{\circ}}
\newcommand{\diskkk}[1]{\accentset{\smash{\raisebox{-0.12ex}{$\scriptstyle\circ$}}}{#1}\rule{0pt}{2.3ex}}
\fboxrule0.0001pt \fboxsep0pt

\section{ Introduction}

These notes present the core of the results of \cite{jk-cromlech}  
with emphasis on the the new algebraic formalism for relativistic addition of velocities.

%

Recall that a loop is a set with binary operation containing a unit element and with left and right division well-defined.  
In short, it is a quasigroup with an identity element(associativity is not assumed).

\section{Menhir loop}


Let $\mathbb F$ be a division algebra, and 
$\diskkk{\mathbb F} = \{ x\in \mathbb F\;\big|\; |x|^2\! <1\}$
be  the unit open disk.
Our standard examples are $\mathbb R$,  $\mathbb C$,  $\mathbb H$, and  $\mathbb O$,  
(real numbers, complex numbers, quaternions and octonions).

\begin{definition}
The {\bf menhir loop} $\mathcal M(\mathbb F)$ is the pair $\{ \disk{\mathbb F},\, \boxplus\}$
where the product is defined by
\begin{equation}
\label{eq:menhir}
a\boxplus b = \frac{a+b}{1+\bar a b}
\end{equation}
and the bar above $a$ denotes the conjugation in $\mathbb F$.
\end{definition}

In case of non-commutative algebra $\mathbb F$,
the ratio \eqref{eq:menhir} is interpreted as   
$a\boxplus b = (a+b)(1+\bar a b)^{-1}$.
The menhir loop has a zero (neutral element):  $0\boxplus a = a\boxplus 0 = a$,
and the negative to $a$ is $-a$.
In general
$\mathcal M(\mathbb F)$ in neither commutative nor associative
unless the field conjugation is trivial (e,.g., when $\mathbb F=\mathbb R$).

\begin{proposition}
\label{thm:identities}
Among the ``soft'' versions of associativity,  the following hold for $\mathcal M(\mathbb F)$
for each tof the algebras $\mathbb F = \mathbb R,\, \mathbb C,\,\mathbb H, \, \mathbb O$:
$$
\begin{array}{lll}
(i)& (a\boxplus a)\boxplus a =  a\boxplus (a\boxplus a ) &\hbox{(power associativity)}  \\
(ii)& (a\boxplus a)\boxplus b =  a\boxplus (a\boxplus b )  &\hbox{(left alternate associativity)}\\
(iii)& a\boxplus (b\boxplus (a\boxplus c)) =  (a\boxplus (b\boxplus a)) \boxplus c \ \   &\hbox{(no name)}
\end{array}
$$
\noindent
{\bf Proof:} Direct inspection.  See Appendix for a diagrammatic system of enumerating potential associativity rules.
\QED
\end{proposition}

In general,  the right alternate associativity does not hold, unless $\mathbb F = \mathbb R$. 
Equality ({\it iii}\/) is the only four-term associativity rule that holds. 
In particular, none of the Moufang identities holds in general in the menhir loop.  
Clearly, the power associativity (i) is a special case of (ii). 
\\

Define  a ``box multiplication by 2'' (doubling) as the map
$$
a \ \ \mapsto \ \ 2\boxdot a  \equiv   a\boxplus a = \frac{2a}{1+|a|^2}
$$
%
which effectively is a non-linear rescaling.  Similarly, define  a ``box-halving'' as:
$$
\frac{1}{2}\boxdot a = \frac{a}{1+\sqrt{1-|a|^2}}
$$
It is easy to check that it is an inverse operation to the doubling:
$$
2\boxdot (\frac{1}{2}\boxdot a) \ = \ \left( 2\cdot \frac{1}{2}\right)\boxdot a \ = \  a\,,
$$
Similarly, $\frac{1}{2}\boxdot(2\boxdot a) = a$.
We may also use the double line fraction to indicate this operation:
$$
                \Dfrac{\,a\,}{2} \  = \   \frac{1}{2}\boxdot a 
$$
Although these definition mimic the analogous operations for the real numbers, 
we do not have the distributivity of the box-addition with respect to box-scaling.
This opens an opportunity for various simple deformations of the loop product.

\begin{definition}
The {\bf double loop} of the menhir loop is a pair 
$\mathcal M_2(\mathbb F)=\{\, \diskkk{\mathbb F},\, \oplus\,\}$ 
were the new product is defined as
\begin{equation}
\label{eq:halfadd}
a\oplus b =  2 \boxdot \left({\Dfrac{a}{2}} \ \boxplus \ \Dfrac{b}{2} \right)
\end{equation}
\end{definition}

Changing somewhat notation to
$$
\begin{array}{lll}
\mu: &\diskkk {\mathbb F} \ \to \  \diskkk{\mathbb F}:           &\ a\mapsto \frac{1}{2}\boxdot a \\
\mu^{-1}: &\diskkk {\mathbb F} \ \to \ \diskkk {\mathbb F}:  &\ a\mapsto 2\boxdot a 
\end{array}
$$
emphasizes that $\mu$ sets an isomorphism 
$\{\,\diskkk{\mathbb F},\, \boxplus\,\} \to \{\,\diskkk{\mathbb F}, \,\oplus\,\}$ 
between the two loops: 
$$
\mu(a\oplus b)= \mu(a) \boxplus \mu(b)
$$
In particular, $\mu(0) = 0$ and $\mu(-a) = -\mu(a)$.
One may view the product $\oplus$ as a {\bf deformation} of $\boxplus$.

%
%
%

%

\section{Relativistic composition of velocities} 
 
The relativistic composition of velocities (called also ``addition''), 
has the following formula (presented first by M{\o}ller (1952)):

\begin{equation}
\label{eq:Moller}
\mathbf v\;\hat \oplus \; \mathbf u   
\ = \frac{    \sqrt{1-|\mathbf v|^2} \, \mathbf u  +
                    \left( \frac{ \left( 1-\sqrt{1-|\mathbf v|^2} \right)\, \mathbf v\cdot \mathbf u }
                                {|\mathbf v|^2}
                                                        +1 \right)\, \mathbf v
}
      {1 + \mathbf v\cdot \mathbf u}
\end{equation}
where 
$\mathbf u$,  $\mathbf v$ and $\mathbf v  \hat \oplus \mathbf v$ 
are vectors of an Euclidean space $\mathbf E$ 
(usually $\mathbf E\cong \mathbb R^3$). 
Finding it rather discouraging, the authors of textbooks on relativity rarely evoke this formula.
It has been noticed that \eqref{eq:Moller} defines technically a loop due to non-associativity of this product, 
see \cite{Sb}.  This feature is highly non-intuitive from the standard, Galilean-based, intuition about bodies in motion.

Here is the remedy:
We can achieve the same result in a much simpler, elegant, 
and transparent way by using the menhir algebra.
It turns out that the 2-deformation loop $\mathcal M_2\mathbb F$ is the loop 
representing the relativistic addition of velocities.


\begin{proposition}
The standard relativistic addition \eqref{eq:Moller} coincides with the menhir altered product  \eqref{eq:halfadd},
$\mathbf v\,\hat \oplus \, \mathbf u  =  v\; \oplus \; u$,
where the non-bold letters denote the vectors re-interpreted as the elements of 
one of the division algebras $\mathbb F$.  
Using only bold letters for clarity:
%
\begin{equation}
\label{eq:iso}
\boxed{
\qquad
~\mathbf v \oplus \mathbf w \ = \
\mu^{-1} \left( \phantom{\big|}\mu(\mathbf v)\boxplus \mu(\mathbf u) \; \right) \phantom{\Big|} \qquad
}
\end{equation}
In particular, the cases of  \,$\mathbb F = \mathbb R$, $\mathbb C$ and  $\mathbb H$
correspond to 1-, 2- and 4-dimensional space models, respectively.
\end{proposition}

\noindent
{\bf Proof:}  One may attempt the rather unpleasantly involved direct algebraic transformations
of the right-hand side of \eqref{eq:iso}  
to the vector expression of \eqref{eq:Moller}.
Alternatively, the derivation of this result from the fundamental principles may be found in \cite {jk-cromlech}.
\QED
\\

The computation of the composition of only two velocities 
may be reduced to the two-dimensional subspace spanned by them.
In such a case complex numbers suffice as the ambient algebraic structure.
The relation between velocities and menhirs  
$\mu (a \voplus b) = \mu(a)\moplus \mu (b)$ is then represented by the following commutative diagram:
\begin{equation}
\label{eq:diagram}
\begin{tikzpicture}[baseline=-0.7ex]
    \matrix (m) [ matrix of math nodes,
                         row sep=1.5em,
                         column sep=4em,
                         text height=3ex, text depth=2ex] 
 {
  \hbox{\sf velocities:} &\quad \mathbb C \times \mathbb C \quad    & \quad \mathbb C  \quad   \\
  \hbox{\sf menhirs:} &\quad  \mathbb C\times \mathbb C  \quad     & \quad \mathbb C \quad    \\
  };
    \path[-stealth]
       (m-1-2) edge [transform canvas={yshift=0.5ex}] node[above] {$\voplus$} (m-1-3)
        (m-2-2) edge node[above] {$\moplus$} (m-2-3)
        (m-1-2) edge node[right]  {$\mu\times\mu$} (m-2-2)
        (m-1-3) edge node[right] {$\mu$} (m-2-3)
;
\end{tikzpicture}   
\end{equation}

This hidden structure of the loop of relativistic addition simplifies enormously the calculations. 
To be explicit: 
Say $a, b\in \mathbb C$ represent velocities.
To obtain their relativistic composition $c=a \oplus b$
we first re-scale them using $\mu$:
$$
\mu: a \mapsto a' = \Dfrac{\,a\,}{2} = \frac{a}{1+\sqrt{1-|a|^2}}\,,
\qquad
\mu: b \mapsto b' = \Dfrac{\,b\,}{2} =  \frac{b}{1+\sqrt{1-|b|^2}}
$$
and add them via the menhir loop product
\begin{equation}
\label{eq:short1}
c' =  a'\boxplus b' = \frac{a'+b'}{1+\bar a' b'}\,.
\end{equation}
Then we scale back with the inverse map, i.e., 
\begin{equation}
\label{eq:short2}
c =  \frac{2c'}{1+|c'|^2}
\end{equation}
and obtain the relativistic composition  $c=a\oplus b$.
\\

\noindent
{\bf Example:}
Here is a numerical example to illustrate the simplicity of the method
for two non-collinear velocities $a$ and $b$:
$$
\begin{array}{cll} 
a=\frac{3}{5}          \qquad&\mathop{\Longrightarrow}\limits^{\mu} \qquad a' = \frac {1}{3} \,,\\[7 pt]
b=\frac{1+2i}{3}  \qquad&\mathop{\Longrightarrow}\limits^\mu \qquad b' = \frac {1+2i}{5} \,,\\
\end{array}
$$
hence
$$
a'\boxplus b'  
= \frac{7+4i}{13}  \qquad\mathop{\Longrightarrow}\limits^{\mu^{-1}} \qquad a\oplus b = \frac {7+4i}{9}
$$ 
The reader may check that this result coincides with one calculated from the standard M{\o}ller's equation \eqref{eq:Moller}.
\\

For composition of a greater number of non-collinear velocities (assuming three-dimensionality of space), 
the algebra of quaternions suffices as the ambient algebraic structure.  
The nonassociativity marks both loops discussed above.

\section{Coda}

{\bf Intriguing resemblance.} 
In the case of the collinear velocities (or in one-dimensional case), 
the relativistic addition formula \eqref{eq:Moller} simplifies to the Poincar\'e formula \cite{Po}, namely 
\begin{equation}
\label{eq:Poincare}
a\oplus b = \frac{a+b}{1+ab}
\end{equation}
Note how complex is the nonlinear case \eqref{eq:Moller} in comparison.

It is remarkable that 
once the velocities are rescaled by $\mu$ to menhirs, the formula becomes similar to the Poincar\'e formula \cite{Po}.
The only difference lies in the conjugation of the first entry in the denominator.
It is a rather mysterious feature that perhaps needs some insight.
\\

\noindent
{\bf Generalizations.}
Other intriguing mathematical structures that directly generalize the standard relativistic case follow naturally.
For instance the division algebra $\mathbb F$ may be replaced by other algebras with conjugation, for instance hyperbolic numbers, octonions, Clifford algebras, etc.
One may also generalize the scaling by 2 a more general map $\mathbb Z \times \mathbb F \to \mathbb F$:
$$
k\boxdot a = a\boxplus a \boxplus \cdots .\boxplus a    \qquad \hbox{($k$ times)}
$$
Similarly, we define scaling by $\frac{1}{k}$ as the inverse operation.
The $k$-loop deformation generalizes to
$$
a\boxplus_k b =  k \boxdot \left({\Dfrac{a}{k}} \ \boxplus \ \Dfrac{b}{k} \right)
$$   

\begin{definition}
A $k$-deformation of the menhir loop is  $\mathcal M_k(\mathbb F) = \{\,\mathbb F, \,\boxplus_k\,\}$
\end{definition}

By the very definition, we have a family of isomorphic structures.
Such cases of $k$-relativity in which 2 is replaced by number $k=3,4,...$  etc., 
are open to interpretations, including the limit  $a\boxplus_\infty b$.

$$
\begin{array}{ccccc}
 k=1  &  &  k=2 & &  k=\infty \\[3pt]
\hbox{\sf Menhir}  &  &  \hbox{\sf Relativistic} & &   \hbox{\sf ?}  \\[-2pt]
\hbox{\sf loop}  &\longleftrightarrow& \hbox{\sf loop} &\longrightarrow& \hbox{\sf ...}  \\[7pt]
\mathcal M(\mathbb F)=\{\diskkk{\mathbb F},\boxplus\}      
               &&\  \mathcal M_2(\mathbb F)=\{\diskkk{\mathbb F},\boxplus_2\equiv \oplus\}    \  
                           &&  \mathcal M_\infty (\mathbb F)=\{\diskkk{\mathbb F},\boxplus_\infty\}\\
\end{array}
$$ 

\newpage

\section*{Appendix:  Diagrammatic of Moufang-like identities}

Here is a simple visualization of bracketing  proposed
for non-associative algebras.  
The figures are rather self-explanatory, see Figure \ref{fig:gorki}:  
\noindent

\begin{figure}[h]
\centering
\includegraphics[scale=.81]{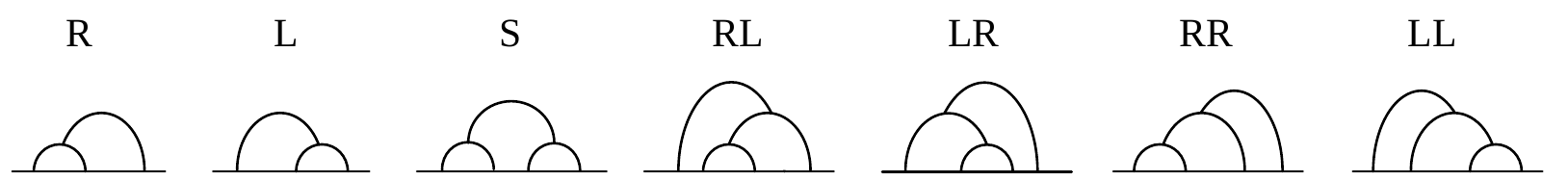}
\caption{\small Four elements bracketed.}
\label{fig:gorki}
\end{figure}

\noindent
If the points on the line are labeled alphabetically, $(a,b,c,...)$, 
then  $R=(ab)c$, \; $L=a(bc)$,\; $S=(ab)(cd)$,\;  $LR=(a(bc))d$, etc.
In order to indicate a repetition an element, the corresponding spot is marked be an open circle.
Here are examples of possible identities:

\begin{figure}[h]
\centering
\includegraphics[scale=.81]{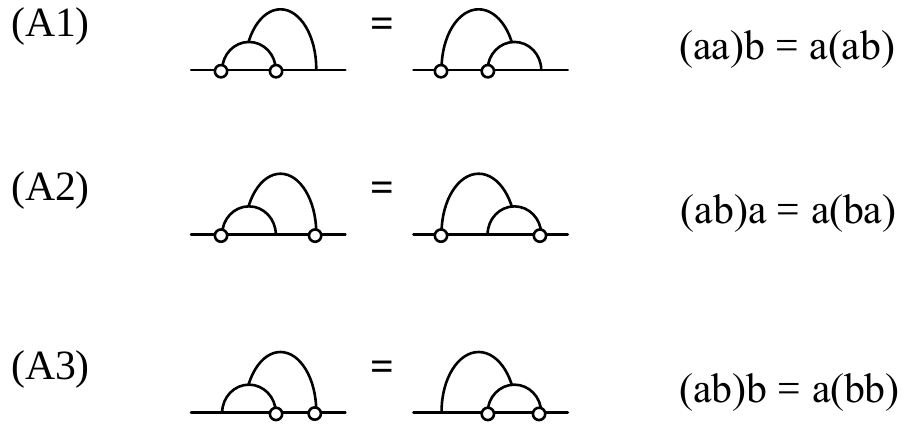}
\caption{\small Alternating associativities.}
\label{fig:a}
\end{figure}

\noindent
Below, the Moufang identities and the unnamed identity of Proposition \ref{thm:identities} (iii)
are presented diagrammatically:

\begin{figure}[h]
\centering
\includegraphics[scale=.81]{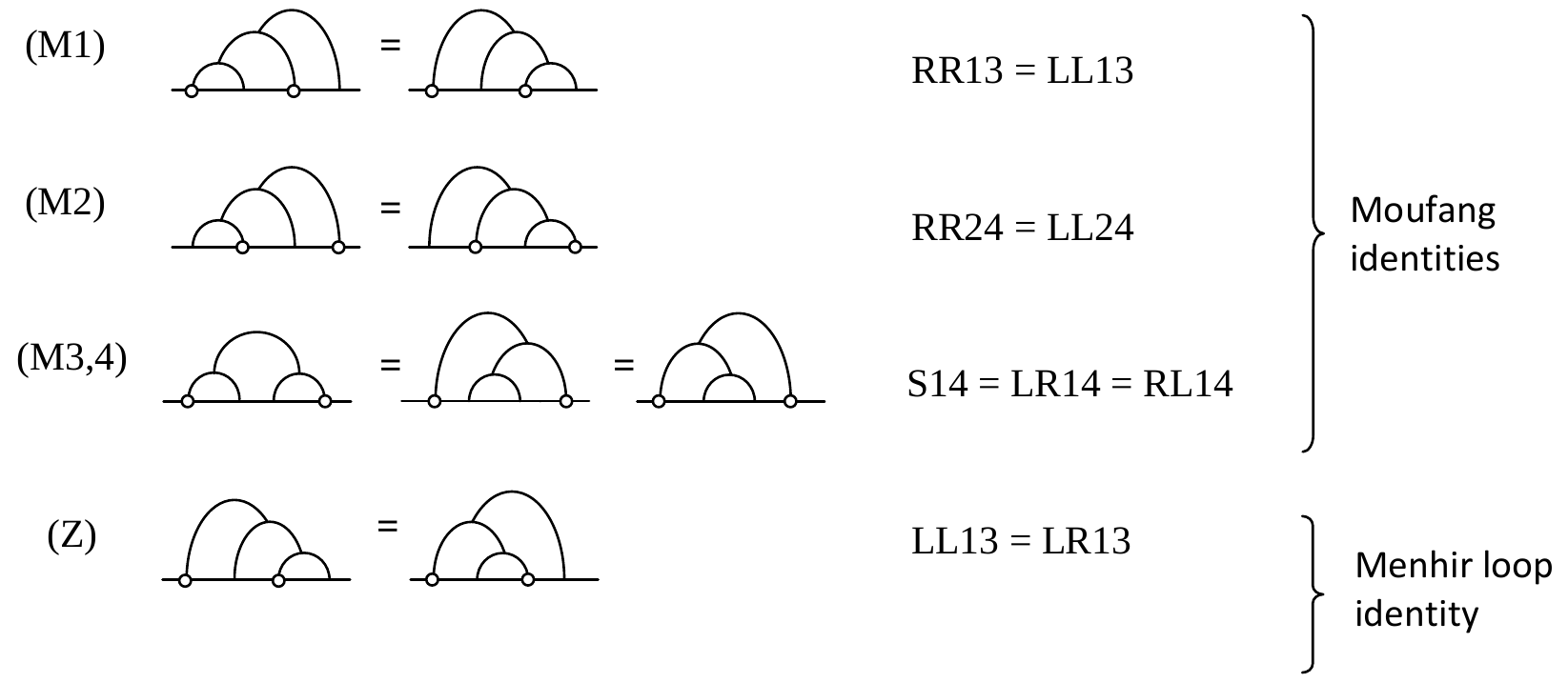}
\caption{\small Some four-elements identities.}
\label{fig:Stonehenge}
\end{figure}

%
%
%

\newpage
\section*{Acknowledgments:}
The author would like to express his gratitude to Zbigniew Oziewicz who 
has called his attention to C. M{\o}ller's book    
and has been always available to discuss these matters
and generously share his thoughts.

%
%


\newpage

\end{document}